\documentclass[11pt]{article}
\usepackage{amssymb,amsmath,amsfonts}
\usepackage{graphicx}
\usepackage{graphics}
\usepackage{eepic,epsfig}

\begin{document}

\title{Topological Casimir effect in compactified cosmic string spacetime }
\author{E. R. Bezerra de Mello$^{1}$\thanks{%
E-mail: emello@fisica.ufpb.br},\, A. A. Saharian$^{1,2}$\thanks{%
E-mail: saharian@ysu.am} \\
\\
\textit{$^{1}$Departamento de F\'{\i}sica, Universidade Federal da Para\'{\i}%
ba}\\
\textit{58.059-970, Caixa Postal 5.008, Jo\~{a}o Pessoa, PB, Brazil}\vspace{%
0.3cm}\\
\textit{$^2$Department of Physics, Yerevan State University,}\\
\textit{1 Alex Manoogian Street, 0025 Yerevan, Armenia}}
\maketitle

\begin{abstract}
We investigate the Wightman function, the vacuum expectation values of the
field squared and the energy-momentum tensor for a massive scalar field with
general curvature coupling in the generalized cosmic string geometry with a
compact dimension along its axis. The boundary condition along the
compactified dimension is taken in general form with an arbitrary phase. The
vacuum expectation values are decomposed into two parts. The first one
corresponds to the uncompactified cosmic string geometry and the second one
is the correction induced by the compactification. The asymptotic behavior
of the vacuum expectation values of the field squared, energy density and
stresses are investigated near the string and at large distances. We show
that the nontrivial topology due to the cosmic string enhances the vacuum
polarization effects induced by the compactness of spatial dimension for
both the field squared and the vacuum energy density. A simple formula is
given for the part of the integrated topological Casimir energy induced by
the planar angle deficit. The results are generalized for a charged scalar
field in the presence of a constant gauge field. In this case, the vacuum
expectation values are periodic functions of the component of the vector
potential along the compact dimension.
\end{abstract}

\bigskip


\bigskip

\section{Introduction}

Within the framework of grand unified theories, as a result of
vacuum symmetry breaking phase transitions, different types of
topological defects could be produced in the early universe
\cite{Kibb80,Vile94}. In particular, cosmic strings have attracted
considerable attention. Although the recent observational data on
the cosmic microwave background have ruled out cosmic strings as
the primary source for large scale structure formation, they are
still candidates for a variety of interesting physical phenomena
such as the generation of gravitational waves \cite{Damo00}, high
energy cosmic rays \cite{Bhat00}, and gamma ray bursts
\cite{Bere01}. Recently the cosmic strings attracted a renewed
interest partly because a variant of their
formation mechanism is proposed in the framework of brane inflation~\cite%
{Sara02}.

In the simplest theoretical model describing the infinite straight cosmic
string the spacetime is locally flat except on the string where it has a
delta shaped curvature tensor. The corresponding nontrivial topology raises
a number of interesting physical effects. One of these concerns the effect
of a string on the properties of quantum vacuum. Explicit calculations for
vacuum polarization effects in the vicinity of a string have been done for
various fields \cite{Hell86}-\cite{BezeKh06}. Vacuum polarization effects by
the cosmic string carrying a magnetic flux are considered in Ref. \cite%
{Guim94}. Another type of topological quantum effects appears in models with
compact spatial dimensions. The presence of compact dimensions is a key
feature of most high energy theories of fundamental physics, including
supergravity and superstring theories. An interesting application of the
field theoretical models with compact dimensions recently appeared in
nanophysics. The long-wavelength description of the electronic states in
graphene can be formulated in terms of the Dirac-like theory in
three-dimensional spacetime with the Fermi velocity playing the role of
speed of light (see, e.g., \cite{Cast09}). Single-walled carbon nanotubes
are generated by rolling up a graphene sheet to form a cylinder and the
background spacetime for the corresponding Dirac-like theory has topology $%
R^{2}\times S^{1}$. The compactification of spatial dimensions serves to
alter vacuum fluctuations of a quantum field and leads to the Casimir-type
contributions in the vacuum expectation values of physical observables (see
Refs. \cite{Most97,Eliz94} for the topological Casimir effect and its role
in cosmology). In the Kaluza-Klein-type models, the topological Casimir
effect induced by the compactification has been used as a stabilization
mechanism for the size of extra dimensions. The Casimir energy can also
serve as a model for dark energy needed for the explanation of the present
accelerated expansion of the universe. The influence of extra compact
dimensions on the Casimir effect in the classical configuration of two
parallel plates has been recently discussed for the case of a scalar field
\cite{Chen06}, for the electromagnetic field with perfectly conducting
boundary conditions \cite{Popp04}, and for a fermionic field with bag
boundary conditions \cite{Bell09}.

In this paper we shall study the configuration with both types of sources
for the vacuum polarization, namely, a generalized cosmic string spacetime
with a compact spatial dimension along its axis (for combined effects of
topology and boundaries on the quantum vacuum in the geometry of a cosmic
string see \cite{Brev95,Beze06,Beze07}). For a massive scalar field with an
arbitrary curvature coupling parameter, we evaluate the Wightman function,
the vacuum expectation values of the field squared and the energy-momentum
tensor. These expectation values are among the most important quantities
characterizing the vacuum state. Though the corresponding operators are
local, due to the global nature of the vacuum, the vacuum expectation values
carry an important information about the the global properties of the bulk.
In addition, the vacuum expectation value of the energy-momentum tensor acts
as the source of gravity in the semiclassical Einstein equations. It
therefore plays an important role in modelling a self-consistent dynamics
involving the gravitational field. The problem under consideration is also
of separate interest as an example with two different kind of topological
quantum effects, where all calculations can be performed in a closed form.

We have organized the paper as follows. The next section is devoted to the
evaluation of the Wightman function for a massive scalar field in a
generalized cosmic string spacetime with a compact dimension. Quasiperiodic
boundary condition with an arbitrary phase is assumed along the compact
dimension . By using the formula for the Wightman function, in section \ref%
{Sec:phi2} we evaluate the vacuum expectation value of the field squared.
This expectation value is decomposed into two parts: the first one
corresponding to the geometry of a cosmic string without compactification
and the second one being induced by the compactification. The vacuum
expectation value of the energy-momentum tensor is discussed in section \ref%
{Sec:EMT}. Section \ref{Sec:VacEn} is devoted to the investigation of the
part in topological vacuum energy induced by the planar angle deficit.
Finally, the results are summarized and discussed in section \ref{sec:Conc}.

\section{Wightman function}

\label{sec:WightFunc}

We consider a $(D+1)$-dimensional generalized cosmic string spacetime.
Considering the generalized cylindrical coordinates $(x^{1},x^{2},\ldots
,x^{D})=(r,\phi ,z,x^{4},\ldots ,x^{D})$ with the string on the $(D-2)$
-dimensional hypersurface $r=0$, the corresponding geometry is described by
the line element
\begin{equation}
ds^{2}=g_{ik}dx^{i}dx^{k}=dt^{2}-dr^{2}-r^{2}d\phi
^{2}-dz^{2}-\sum_{l=4}^{D}(dx^{l})^{2}.  \label{ds21}
\end{equation}%
The coordinates take values in the ranges $r\geqslant 0$, $0\leqslant \phi
\leqslant \phi _{0}$, $-\infty <x^{l}<+\infty $ for $l=4,\ldots ,D$, and the
spatial points $(r,\phi ,z,x^{4},\ldots ,x^{D})$ and $(r,\phi +\phi
_{0},x^{4},\ldots ,x^{D})$ are to be identified. Additionally we shall
assume that the direction along the $z$-axis is compactified to a circle
with the length $L$: $0\leqslant z\leqslant L$ (about the generalization of
the model in the case of an arbitrary number of compact dimensions along the
axis of the string see below). In the standard $D=3$ cosmic string case with
$-\infty <z<+\infty $, the planar angle deficit is related to the mass per
unit length of the string $\mu $ by $2\pi -\phi _{0}=8\pi G\mu $, where $G$
is the Newton gravitational constant.

It is interesting to note that the effective metric produced in superfluid $%
^{3}\mathrm{He-A}$ by a radial disgyration is described by the $D=3$ line
element (\ref{ds21}) with the negative planar angle deficit \cite{Volo98}.
In this condensed matter system the role of the Planck energy scale is
played by the gap amplitude. The graphitic cones are another class of
condensed matter systems, described in the long wavelength approximation by
the metric (\ref{ds21}) with $D=2$. Graphitic cones are obtained from the
graphene sheet if one or more sectors are excised. The opening angle of the
cone is related to the number of sectors removed, $N_{c}$, by the formula $%
2\pi (1-N_{c}/6)$, with $N_{c}=1,2,\ldots ,5$. All these angles have been
observed in experiments \cite{Kris97}. Induced fermionic current and
fermionic condensate in a $(2+1)$-dimensional conical spacetime in the
presence of a circular boundary and a magnetic flux have been recently
investigated in Ref. \cite{Beze10}.

In this paper we are interested in the calculation of one-loop quantum
vacuum effects for a scalar quantum field $\varphi (x)$, induced by the
non-trivial topology of the $z$-direction in the geometry (\ref{ds21}). For
a massive field with curvature coupling parameter $\xi $ the field equation
has the form
\begin{equation}
\left( \nabla ^{i}\nabla _{i}+m^{2}+\xi R\right) \varphi (x)=0,
\label{fieldeq}
\end{equation}%
where $\nabla _{i}$ is the covariant derivative operator and $R$
is the scalar curvature for the background spacetime. In the
geometry under consideration $R=0$ for $r\neq 0$. The values of
the curvature coupling parameter $\xi =0$ and $\xi =\xi _{D}\equiv
(D-1)/4D$ correspond to the most important special cases of
minimally and conformally coupled scalars, respectively. We assume
that along the compact dimension the field obeys quasiperiodicity
condition
\begin{equation}
\varphi (t,r,\phi ,z+L,x^{4},\ldots ,x^{D})=e^{2\pi i\beta }\varphi
(t,r,\phi ,z,x^{4},\ldots ,x^{D}),  \label{Period}
\end{equation}%
with a constant phase $\beta $, $0\leqslant \beta \leqslant 1$. The special
cases $\beta =0$ and $\beta =1/2$ correspond to the untwisted and twisted
fields, respectively, along the $z$-direction. We could also consider
quasiperiodicity condition with respect to $\phi \rightarrow \phi +\phi _{0}$%
. This would correspond to a cosmic string which carries an internal
magnetic flux. Though the corresponding generalization is straightforward,
for simplicity we shall consider a string without a magnetic flux.

In quantum field theory, the imposition of the condition (\ref{Period})
changes the spectrum of the vacuum fluctuations compared to the case with
uncompactified dimension. As a consequence, the vacuum expectation values
(VEVs) of physical observables are changed. The properties of the vacuum
state are described by the corresponding positive frequency Wightman
function, $W(x,x^{\prime })=\langle 0|\varphi (x)\varphi (x^{\prime
})|0\rangle $, where $|0\rangle $ stands for the vacuum state. In
particular, having this function we can evaluate the VEVs of the field
squared and the energy-momentum tensor. In addition, the response of
particle detectors in an arbitrary state of motion is determined by this
function (see, for instance, \cite{Birr82,Taga86}). For the evaluation of
the Wightman function we use the mode sum formula
\begin{equation}
W(x,x^{\prime })=\sum_{\mathbf{\alpha }}\varphi _{\mathbf{\alpha }%
}(x)\varphi _{\mathbf{\alpha }}^{\ast }(x^{\prime }),  \label{vevWf}
\end{equation}%
where $\{\varphi _{\mathbf{\alpha }}(x),\varphi _{\mathbf{\alpha }}^{\ast
}(x)\}$ is a complete set of normalized mode functions satisfying the
periodicity condition (\ref{Period}) and $\alpha $ is a collective notation
for the quantum numbers specifying the solution.

In the problem under consideration, the mode functions are specified by the
set of quantum numbers $\alpha =(n,\gamma ,k_{z},\mathbf{k})$, with the
values in the ranges $n=0,\pm 1,\pm 2,\ldots $, $\mathbf{k}=(k_{4},\ldots
,k_{D})$, $-\infty <k_{l}<\infty $. The mode functions have the form%
\begin{equation}
\varphi _{\alpha }(x)=\frac{q\gamma J_{q\left\vert n\right\vert }(\gamma r)}{%
2(2\pi )^{D-2}\omega L}\exp \left( iqn\phi +ik_{z}z+i\mathbf{kr}_{\parallel
}-i\omega t\right) ,  \label{Eigfunccirc}
\end{equation}
where $J_{\nu }(z)$ is the Bessel function, $\mathbf{r}_{\parallel
}=(x^{4},\ldots ,x^{D})$ and%
\begin{equation}
\omega =\sqrt{\gamma ^{2}+k_{z}^{2}+k^{2}+m^{2}},\quad q=2\pi /\phi _{0}.
\label{qu}
\end{equation}
From the periodicity condition (\ref{Period}), for the eigenvalues of the
quantum number $k_{z}$ one finds%
\begin{equation}
k_{z}=2\pi (l+\beta )/L,\;l=0,\pm 1,\pm 2,\ldots .  \label{Eigkz}
\end{equation}

Substituting the mode functions (\ref{Eigfunccirc}) into the sum (\ref{vevWf}%
), for the positive frequency Wightman function one finds
\begin{eqnarray}
W(x,x^{\prime }) &=&\frac{q}{(2\pi )^{D-2}L}\sideset{}{'}{\sum}%
_{n=0}^{\infty }\cos (qn\Delta \phi )\int d\mathbf{k}\,e^{i\mathbf{k}\Delta
\mathbf{r}_{\parallel }}  \notag \\
&&\times \int_{0}^{\infty }d\gamma \,\gamma J_{qn}(\gamma r)J_{qn}(\gamma
r^{\prime })\sum_{l=-\infty }^{\infty }\frac{e^{-i\omega \Delta
t+ik_{z}\Delta z}}{\omega },  \label{WF1}
\end{eqnarray}%
where $\Delta \phi =\phi -\phi ^{\prime }$, $\Delta \mathbf{r}_{\parallel }=%
\mathbf{r}_{\parallel }-\mathbf{r}_{\parallel }^{\prime }$, $\Delta
t=t-t^{\prime }$, $\Delta z=z-z^{\prime }$, and the prime on the sum over $n$
means that the summand with $n=0$ should be taken with the weight 1/2. For
the further evaluation, we apply to the series over $l$ the Abel-Plana
summation formula in the form \cite{Bell10} (for generalizations of the
Abel-Plana formula and their applications in quantum field theory see \cite%
{Most97,SahaBook,Saha06Sum})
\begin{eqnarray}
&&\sum_{l=-\infty }^{\infty }g(l+\beta )f(|l+\beta |)=\int_{0}^{\infty }du\,
\left[ g(u)+g(-u)\right] f(u)  \notag \\
&&\qquad +i\int_{0}^{\infty }du\left[ f(iu)-f(-iu)\right] \sum_{\lambda =\pm
1}\frac{g(i\lambda u)}{e^{2\pi (u+i\lambda \beta )}-1}.  \label{sumform}
\end{eqnarray}%
In the special case of $g(y)=1$ and $\beta =0$ this formula is reduced to
the standard Abel-Plana formula. Taking in Eq. (\ref{sumform})
\begin{equation}
g(y)=e^{2\pi iy\Delta z/L},\;f(y)=\frac{e^{-i\Delta t\sqrt{(2\pi
y/L)^{2}+\gamma ^{2}+k^{2}+m^{2}}}}{\sqrt{(2\pi y/L)^{2}+\gamma
^{2}+k^{2}+m^{2}}},  \label{fg}
\end{equation}%
we present the Wightman function in the decomposed form%
\begin{eqnarray}
&&W(x,x^{\prime })=W_{\text{s}}(x,x^{\prime })+\frac{2q}{(2\pi )^{D-1}}%
\sideset{}{'}{\sum}_{n=0}^{\infty }\cos (qn\Delta \phi )\int d\mathbf{k}%
\,e^{i\mathbf{k}\Delta \mathbf{r}_{\parallel }}\int_{0}^{\infty }d\gamma
\,\gamma J_{qn}(\gamma r)  \notag \\
&&\qquad \times J_{qn}(\gamma r^{\prime })\int_{\sqrt{\gamma ^{2}+k^{2}+m^{2}%
}}^{\infty }dy\frac{\cosh (\Delta t\sqrt{y^{2}-\gamma ^{2}-k^{2}-m^{2}})}{%
\sqrt{y^{2}-\gamma ^{2}-k^{2}-m^{2}}}\sum_{\lambda =\pm 1}\frac{e^{-\lambda
y\Delta z}}{e^{Ly+2\pi i\lambda \beta }-1},  \label{WF1b}
\end{eqnarray}%
where $W_{\text{s}}(x,x^{\prime })$ is the Wightman function in the geometry
of a cosmic string without compactification. The latter corresponds to the
first term on the right hand side of (\ref{sumform}). The integration over
the angular part of $\mathbf{k}$ is done with the help of the formula%
\begin{equation}
\int d\mathbf{k}\,e^{i\mathbf{k}\Delta \mathbf{r}_{\parallel }}F(k)=\frac{%
(2\pi )^{(D-3)/2}}{|\Delta \mathbf{r}_{\parallel }|^{(D-5)/2}}%
\int_{0}^{\infty }dk\,k^{(D-3)/2}J_{(D-5)/2}(k|\Delta \mathbf{r}_{\parallel
}|)F(k),  \label{IntForm1}
\end{equation}%
for a given function $F(k)$. By using the expansion $[e^{Ly+2\pi i\lambda
\beta }-1]^{-1}=\sum_{l=1}^{\infty }e^{-lLy-2\pi il\lambda \beta }$, the
further integrations over $k$ and $y$ are done by making use of formulas
from \cite{Prud86}. Similar transformations are done for the part $W_{\text{s%
}}(x,x^{\prime })$. As a result we find the following expression%
\begin{eqnarray}
W(x,x^{\prime }) &=&\frac{2q}{(2\pi )^{(D+1)/2}}\sideset{}{'}{\sum}%
_{n=0}^{\infty }\cos (qn\Delta \phi )\int_{0}^{\infty }d\gamma \,\gamma
J_{qn}(\gamma r)J_{qn}(\gamma r^{\prime })  \notag \\
&&\times \sum_{l=-\infty }^{\infty }e^{-2\pi li\beta }\left( \gamma
^{2}+m^{2}\right) ^{(D-3)/2}f_{(D-3)/2}(w_{l}\sqrt{\gamma ^{2}+m^{2}}),
\label{WF2}
\end{eqnarray}%
where%
\begin{equation}
w_{l}^{2}=\left( \Delta z+lL\right) ^{2}+|\Delta \mathbf{r}_{\parallel
}|^{2}-(\Delta t)^{2},  \label{wl}
\end{equation}%
and we use the notation
\begin{equation}
f_{\nu }(x)=K_{\nu }(x)/x^{\nu }.  \label{fnu}
\end{equation}%
The term $l=0$ in (\ref{WF2}) corresponds to the function $W_{\text{s}%
}(x,x^{\prime })$.

For the further transformation of the expression (\ref{WF2}) we employ the
integral representation of the modified Bessel function \cite{Wats44},%
\begin{equation}
K_{\nu }(y)=\frac{1}{2^{\nu +1}y^{\nu }}\int_{0}^{\infty }d\tau \frac{%
e^{-\tau y^{2}-1/(4\tau )}}{\tau ^{\nu +1}}.  \label{Kint}
\end{equation}%
Substituting this representation in (\ref{WF2}), the integration over $%
\gamma $ is done explicitly. Introducing an integration variable $u=1/(2\tau
)$ and by changing $l\rightarrow -l$, one finds
\begin{equation}
W(x,x^{\prime })=\frac{q}{(2\pi )^{(D+1)/2}}\sum_{l=-\infty }^{\infty
}e^{2\pi li\beta }\int_{0}^{\infty
}du\,u^{(D-3)/2}e^{-u_{l}^{2}u/2-m^{2}/(2u)}\sideset{}{'}{\sum}%
_{n=0}^{\infty }\cos (qn\Delta \phi )I_{qn}(urr^{\prime }),  \label{WF3}
\end{equation}%
where%
\begin{equation}
u_{l}^{2}=r^{2}+r^{\prime 2}+\left( \Delta z-lL\right) ^{2}+|\Delta \mathbf{r%
}_{\parallel }|^{2}-(\Delta t)^{2}.  \label{ul}
\end{equation}

The expression for the Wightman function may be further simplified by using
the formula%
\begin{equation}
\sideset{}{'}{\sum}_{m=0}^{\infty }\cos (qm\Delta \phi )I_{qm}\left(
w\right) =\frac{1}{2q}\sum_{k}e^{w\cos (2k\pi /q-\Delta \phi )}-\frac{1}{%
4\pi }\sum_{j=\pm 1}\int_{0}^{\infty }dy\frac{\sin (q\pi +jq\Delta \phi
)e^{-w\cosh y}}{\cosh (qy)-\cos (q\pi +jq\Delta \phi )},  \label{sumform2}
\end{equation}%
where the summation in the first term on the right hind side goes under the
condition
\begin{equation}
-q/2+q\Delta \phi /(2\pi )\leqslant k\leqslant q/2+q\Delta \phi /(2\pi ).
\label{SumCond}
\end{equation}%
The formula (\ref{sumform2}) is obtained by making use of the integral
representation 9.6.20 from \cite{Abra72} for the modified Bessel function
and changing the order of the summation and integrations. Note that, for
integer values of $q$, formula (\ref{sumform2}) reduces to the well-known
result \cite{Prud86,Spin08}%
\begin{equation}
\sideset{}{'}{\sum}_{m=0}^{\infty }\cos (qm\Delta \phi )I_{qm}\left(
w\right) =\frac{1}{2q}\sum_{k=0}^{q-1}e^{w\cos (2k\pi /q-\Delta \phi )}.
\label{SumFormSp}
\end{equation}%
Substituting (\ref{sumform2}) with $w=urr^{\prime }$ into (\ref{WF3}), the
integration over $u$ is performed explicitly in terms of the modified Bessel
function and one finds
\begin{eqnarray}
W(x,x^{\prime }) &=&\frac{m^{D-1}}{(2\pi )^{(D+1)/2}}\sum_{l=-\infty
}^{\infty }e^{2\pi li\beta }\Bigg[\sum_{k}f_{(D-1)/2}(m\sqrt{%
u_{l}^{2}-2rr^{\prime }\cos (2\pi k/q-\Delta \phi )})  \notag \\
&&-\frac{q}{2\pi }\sum_{j=\pm 1}\sin (q\pi +jq\Delta \phi )\int_{0}^{\infty
}dy\frac{f_{(D-1)/2}(m\sqrt{u_{l}^{2}+2rr^{\prime }\cosh y})}{\cosh
(qy)-\cos (q\pi +jq\Delta \phi )}\Bigg],  \label{WF4}
\end{eqnarray}%
with the notation from (\ref{fnu}). This is the final expression of the
Wightman function for the evaluation of the VEVs in the following sections.
It allows us to present the VEVs of the field squared and the
energy-momentum tensor for a scalar massive field in a closed form for
general value of $q$. In the special case of integer $q$, the general
formula is reduced to%
\begin{equation}
W(x,x^{\prime })=\frac{m^{D-1}}{(2\pi )^{(D+1)/2}}\sum_{l=-\infty }^{\infty
}e^{2\pi li\beta }\sum_{k=0}^{q-1}f_{(D-1)/2}(m\sqrt{u_{l}^{2}-2rr^{\prime
}\cos (2\pi k/q-\Delta \phi )}).  \label{WF4Sp}
\end{equation}%
In this case the Wightman function is expressed in terms of the $q$ images
of the Minkowski spacetime function with a compactified dimension along the
axis $z$.

For a massless field, from (\ref{WF4}) one finds%
\begin{eqnarray}
W(x,x^{\prime }) &=&\frac{\Gamma ((D-1)/2)}{4\pi ^{(D+1)/2}}\sum_{l=-\infty
}^{\infty }e^{2\pi li\beta }\bigg[ \sum_{k}(u_{l}^{2}-2rr^{\prime }\cos
(2\pi k/q-\Delta \phi ))^{(1-D)/2}  \notag \\
&& -\frac{q}{2\pi }\sum_{j=\pm 1}\sin (q\pi +jq\Delta \phi )\int_{0}^{\infty
}dy\frac{(u_{l}^{2}+2rr^{\prime }\cosh y)^{(1-D)/2}}{\cosh (qy)-\cos (q\pi
+jq\Delta \phi )}\bigg] .  \label{WF4m0}
\end{eqnarray}%
The $l=0$ term in the expressions above corresponds to the Wightman function
in the geometry of a cosmic string without compactification:
\begin{eqnarray}
W_{\text{s}}(x,x^{\prime }) &=&\frac{m^{D-1}}{(2\pi )^{(D+1)/2}}\bigg[ %
\sum_{k}f_{(D-1)/2}(m\sqrt{u_{0}^{2}-2rr^{\prime }\cos (2\pi k/q-\Delta \phi
)})  \notag \\
&& -\frac{q}{2\pi }\sum_{j=\pm 1}\sin (q\pi +jq\Delta \phi )\int_{0}^{\infty
}dy\frac{f_{(D-1)/2}(m\sqrt{u_{0}^{2}+2rr^{\prime }\cosh y})}{\cosh
(qy)-\cos (q\pi +jq\Delta \phi )}\bigg] ,  \label{WF4Unc}
\end{eqnarray}%
where $u_{0}^{2}$ is given by (\ref{ul}) with $l=0$.

The formulas given above can be generalized for a charged scalar field $%
\varphi (x)$ in the presence of a gauge field with the vector potential $%
A_{l}=$const and $A_{l}=0$ for $l=0,1,2$. Though the corresponding magnetic
field strength vanishes, the nontrivial topology of the background spacetime
leads to Aharonov-Bohm-like effects for the VEVs. By the gauge
transformation $A_{l}=A_{l}^{\prime }+\partial _{l}\Lambda (x)$, $\varphi
(x)=\varphi ^{\prime }(x)e^{-ie\Lambda (x)}$, with the function $\Lambda
(x)=A_{l}x^{l}$, we can see that the new field $\varphi ^{\prime }(x)$
satisfies the field equation with $A_{l}^{\prime }=0$ and the
quasiperiodicity conditions similar to~(\ref{Period}): $\varphi ^{\prime
}(t,r,\phi ,z+L,x^{4},\ldots ,x^{D})=e^{2\pi i\beta ^{\prime }}\varphi
^{\prime }(t,r,\phi ,z,x^{4},\ldots ,x^{D})$, with%
\begin{equation}
\beta ^{\prime }=\beta +eA_{3}L/(2\pi ).  \label{beta}
\end{equation}%
Hence, for a charged scalar field the corresponding expression for the
Wightman function is obtained from (\ref{WF4}) by the replacement $\beta
\rightarrow \beta ^{\prime }$. In this case the VEVs are periodic functions
of the component of the vector potential along the compact dimension.

We can consider a more general class of compactifications having the spatial
topology $(S^{1})^{p}$ with compact dimensions $(x^{3}=z,x^{4},\ldots
,x^{p}) $, $p\leqslant D$. The phases in the quasiperiodicity conditions
along separate dimensions can be different. For the eigenvalues of the
quantum numbers $k_{i}$, $i=3,\ldots ,p$, one has $2\pi (l_{i}+\beta
_{i})/L_{i}$, $l_{i}=0,\pm 1,\ldots $, with $L_{i}$ being the length of the
compact dimension along the axis $x^{i}$. The mode sum for the corresponding
Wightman function contains the summation over $l_{i}$, $i=3,\ldots ,p$, and
the integration over $k_{i}$ with $i=p+1,\ldots ,D$. We apply to the series
over $l_{p}$ the formula (\ref{sumform}). The term in the expression of the
Wightman function which corresponds to the first integral on the right of (%
\ref{sumform}) is the Wightman function for the topology $(S^{1})^{p-1}$
with compact dimensions $(x^{3}=z,x^{4},\ldots ,x^{p-1})$, and the second
term gives the part induced by the compactness of the direction $x^{p}$. As
a result a recurrence formula is obtained which relates the Wightmann
functions for the topologies $(S^{1})^{p}$ and $(S^{1})^{p-1}$.

The formulas for the Wightman function, given in this section, can be used
to study the response of the Unruh-DeWitt type particle detector (see \cite%
{Birr82,Taga86}) moving in the region outside the string. This response in
the standard geometry of a $D=3$ cosmic string with integer values of the
parameter $q$ has been investigated in \cite{Davi88}. Our main interest in
this paper are the VEVs of the field squared and the energy-momentum tensor
and we turn to the evaluation of these quantities.

\section{VEV of the field squared}

\label{Sec:phi2}

The VEV of the field squared is obtained from the Wightman function by
taking the coincidence limit of the arguments. It is presented in the
decomposed form:%
\begin{equation}
\langle \varphi ^{2}\rangle =\langle \varphi ^{2}\rangle _{\text{s}}+\langle
\varphi ^{2}\rangle _{\text{t}},  \label{VEV2dec}
\end{equation}%
where $\langle \varphi ^{2}\rangle _{\text{s}}$ is the corresponding VEV in
the geometry of a string without compact dimensions and $\langle \varphi
^{2}\rangle _{\text{t}}$ is the topological part induced by the
compactification of the $z$-direction. Because the compactification does not
change the local geometry of the cosmic string spacetime, the divergences in
the coincidence limit are contained in the term $\langle \varphi ^{2}\rangle
_{\text{s}}$ only and the topological part is finite. For $r\neq 0$ the
renormalization of $\langle \varphi ^{2}\rangle _{\text{s}}$ is reduced to
the subtraction of the corresponding quantity in the Minkowski spacetime:
\begin{equation}
\langle \varphi ^{2}\rangle _{\text{s}}=\lim_{x^{\prime }\rightarrow x}[W_{%
\text{s}}(x,x^{\prime })-W_{\text{M}}(x,x^{\prime })],  \label{ph2sren}
\end{equation}%
with $W_{\text{M}}(x,x^{\prime })$ being the Wightman function in the
Minkowski spacetime. The latter coincides with the $k=0$ term in the square
brackets of (\ref{WF4Unc}). The subtraction of the Minkowskian Wightman
function in (\ref{ph2sren}) removes the pole. As a result, one finds%
\begin{equation}
\langle \varphi ^{2}\rangle _{\text{s}}=\frac{2m^{D-1}}{(2\pi )^{(D+1)/2}}%
\left[ \sum_{k=1}^{[q/2]}f_{(D-1)/2}(2mrs_{k})-\frac{q}{\pi }\sin (q\pi
)\int_{0}^{\infty }dy\frac{f_{(D-1)/2}(2mr\cosh (y))}{\cosh (2qy)-\cos (q\pi
)}\right] ,  \label{phi2s}
\end{equation}%
where $[q/2]$ means the integer part of $q/2$ and we have defined%
\begin{equation}
s_{k}=\sin (\pi k/q).  \label{sk}
\end{equation}%
For $1\leqslant q<2$ the first term in the square brackets is absent. The
VEV given by (\ref{phi2s}) is positive for $q>1$.

For a massless field, from (\ref{phi2s}) we obtain the expression below:
\begin{equation}
\langle \varphi ^{2}\rangle _{\text{s}}=\frac{2\Gamma ((D-1)/2)}{(4\pi
)^{(D+1)/2}r^{D-1}}g_{D}(q),  \label{phi2sm0}
\end{equation}%
with the function $g_{D}(q)$ defined as%
\begin{equation}
g_{D}(q)=\sum_{k=1}^{[q/2]}s_{k}^{1-D}-\frac{q}{\pi }\sin (q\pi
)\int_{0}^{\infty }dy\frac{\cosh ^{1-D}(y)}{\cosh (2qy)-\cos (q\pi )}.
\label{gD}
\end{equation}%
The latter is a monotonic increasing positive function of $q$ for $q>1$. For
large values $q$, the dominant contribution to $g_{D}(q)$ comes from the
first term in the square brackets on the right hand side and one finds%
\begin{equation}
g_{D}(q)\approx \zeta (D-1)(q/\pi )^{D-1},\;q\gg 1,  \label{gDlargeq}
\end{equation}%
with $\zeta (x)$ being the Riemann zeta function. Simple expressions for $%
g_{D}(q)$ can be found for odd values of spatial dimension. In particular,
for $D=3,5$ one has
\begin{equation}
g_{3}(q)=\frac{q^{2}-1}{6},\;g_{5}(q)=\frac{\left( q^{2}-1\right) \left(
q^{2}+11\right) }{90}.  \label{g35}
\end{equation}%
The expressions for higher odd values of $D$ can be obtained by using the
recurrence scheme described in \cite{Beze06}. From (\ref{phi2sm0}) we obtain
the results previously derived in \cite{Sour92} for the case $D=2$ and in
\cite{Line87,Smit89} for $D=3$. For a massive field, the leading term in the
asymptotic expansion of the field squared for points near the string, $mr\ll
1$, coincides with (\ref{phi2sm0}). At large distances from the string, $%
mr\gg 1$, the VEV (\ref{phi2s}) is exponentially suppressed.

For the topological part, from (\ref{WF4}) we directly obtain%
\begin{eqnarray}
\langle \varphi ^{2}\rangle _{\text{t}} &=&\frac{4m^{D-1}}{(2\pi )^{(D+1)/2}}%
\sum_{l=1}^{\infty }\cos (2\pi l\beta )\left[ \sideset{}{'}{\sum}%
_{k=0}^{[q/2]}f_{(D-1)/2}(m\sqrt{4r^{2}s_{k}^{2}+\left( lL\right) ^{2}}%
)\right.  \notag \\
&&\left. -\frac{q}{\pi }\sin (q\pi )\int_{0}^{\infty }dy\frac{f_{(D-1)/2}(m%
\sqrt{4r^{2}\cosh ^{2}(y)+\left( lL\right) ^{2}})}{\cosh (2qy)-\cos (q\pi )}%
\right] ,  \label{phi2t}
\end{eqnarray}%
where the prime on the sign of the summation means that the $k=0$ term
should be halved. Note that topological part is not changed under the
replacement $\beta \rightarrow 1-\beta $. An alternative form of the VEV is
obtained from (\ref{WF3}):%
\begin{equation}
\langle \varphi ^{2}\rangle _{\text{t}}=\frac{2qr^{1-D}}{(2\pi )^{(D+1)/2}}%
\sum_{l=1}^{\infty }\cos (2\pi l\beta )\int_{0}^{\infty
}dy\,y^{(D-3)/2}e^{-(2y+\left( lL/r\right) ^{2}y+m^{2}r^{2}/y)/2}%
\sideset{}{'}{\sum}_{n=0}^{\infty }I_{qn}(y).  \label{phi2tb}
\end{equation}%
For a massless field the integral in this formula is expressed in terms of
the associated Legendre function. In particular, from (\ref{phi2tb}) it
follows that the topological part is always positive for an untwisted scalar
($\beta =0$) and it is always negative for a twisted scalar ($\beta =1/2$).
In both cases, $|\langle \varphi ^{2}\rangle _{\text{t}}|$ is a
monotonically decreasing function of the field mass. In the special case of
integer $q$, the general formula (\ref{phi2t}) is reduced to%
\begin{equation}
\langle \varphi ^{2}\rangle _{\text{t}}=\frac{2m^{D-1}}{(2\pi )^{(D+1)/2}}%
\sum_{k=0}^{q-1}\sum_{l=1}^{\infty }\cos (2\pi l\beta )f_{(D-1)/2}(m\sqrt{%
4r^{2}s_{k}^{2}+\left( lL\right) ^{2}}).  \label{phi2tsp}
\end{equation}%
In the discussion below, we shall be mainly concerned with the topological
part. In the presence of a constant gauge field the corresponding
expressions for the VEV of the field squared are obtained by the replacement
$\beta \rightarrow \beta ^{\prime }$ with $\beta ^{\prime }$ defined by (\ref%
{beta}).

Unlike the pure string part, $\langle \varphi ^{2}\rangle _{\text{s}}$, the
topological part is finite on the string. Putting in (\ref{phi2t}) $r=0$ and
using the relation%
\begin{equation}
\int_{0}^{\infty }dy\frac{\sin (q\pi )}{\cosh (qy)-\cos (q\pi )}=\left\{
\begin{array}{cc}
\pi (1-\delta )/q, & q=2p_{0}+\delta , \\
\pi /q, & q=2p_{0}%
\end{array}%
\right. ,  \label{IntForm}
\end{equation}%
with $\delta $ defined in accordance with $q=2p_{0}+\delta $,$\;0\leqslant
\delta <2$, and $p_{0}$ being an integer, one finds%
\begin{equation}
\langle \varphi ^{2}\rangle _{\text{t},r=0}=q\langle \varphi ^{2}\rangle _{%
\text{t}}^{\text{(M)}}=\frac{2qm^{D-1}}{(2\pi )^{(D+1)/2}}\sum_{l=1}^{\infty
}\cos (2\pi l\beta )f_{(D-1)/2}(lmL).  \label{phi2r0}
\end{equation}%
Here $\langle \varphi ^{2}\rangle _{\text{t}}^{\text{(M)}}$ is the VEV of
the field squared in the Minkowski spacetime with a compact dimension of the
length $L$. Note that in the case of $\delta =0$ the left hand side of (\ref%
{IntForm}) is understood in the sense of the limit $\delta \rightarrow 0$.
For points near the string, $mr\ll 1$, the pure stringy part behaves as $%
1/r^{D-1}$ and for $r\ll L$ it dominates in the total VEV.

For a massless field one finds%
\begin{eqnarray}
\langle \varphi ^{2}\rangle _{\text{t}} &=&\frac{\Gamma ((D-1)/2)}{\pi
^{(D+1)/2}L^{D-1}}\sum_{l=1}^{\infty }\cos (2\pi l\beta )\bigg[%
\sideset{}{'}{\sum}_{k=0}^{[q/2]}[(2rs_{k}/L)^{2}+l^{2}]^{(1-D)/2}  \notag \\
&&-\frac{q}{\pi }\sin (q\pi )\int_{0}^{\infty }dy\frac{[(2r/L)^{2}\cosh
^{2}(y)+l^{2}]^{(1-D)/2}}{\cosh (2qy)-\cos (q\pi )}\bigg].  \label{phi2tm0}
\end{eqnarray}%
At small distances from the string, $r\ll L$, in the leading order we have $%
\langle \varphi ^{2}\rangle _{\text{t}}\approx q\langle \varphi ^{2}\rangle
_{\text{t}}^{\text{(M)}}$. At large distances, $r\gg L$, the dominant
contribution to (\ref{phi2tm0}) comes from the $k=0$ term and one has $%
\langle \varphi ^{2}\rangle _{\text{t}}\approx \langle \varphi ^{2}\rangle _{%
\text{t}}^{\text{(M)}}$. As it can be seen from (\ref{phi2t}), we have the
same asymptotics in the case of a massive field as well.

\begin{figure}[tbph]
\begin{center}
\begin{tabular}{cc}
\epsfig{figure=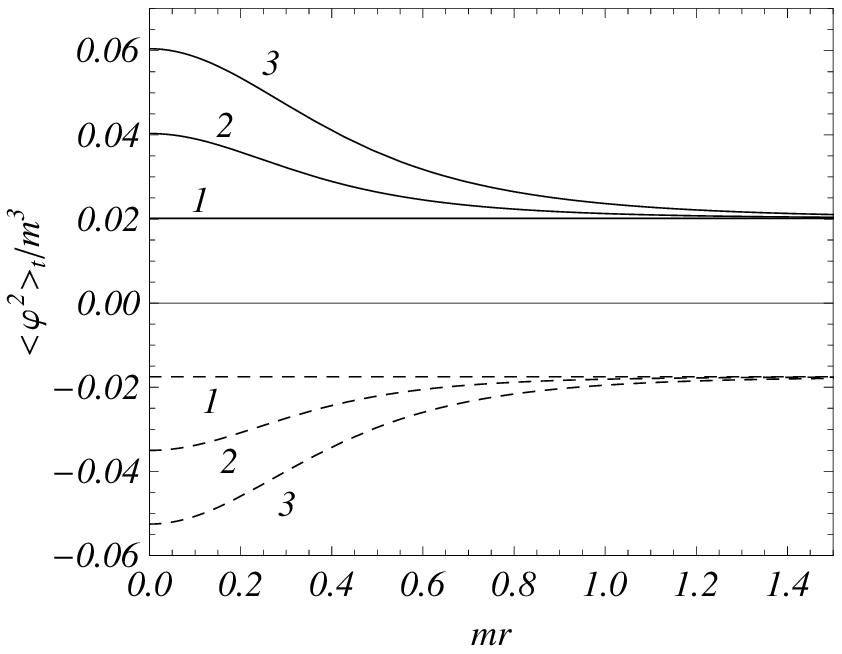, width=7.cm, height=6.cm} & \quad %
\epsfig{figure=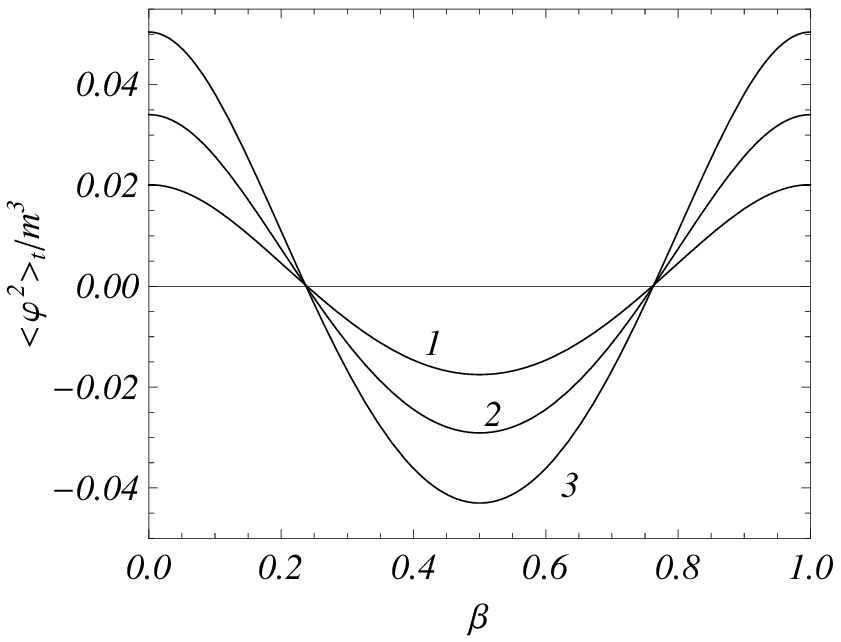, width=7.cm, height=6.cm}%
\end{tabular}%
\end{center}
\caption{The topological part of the VEV in the field squared, $\langle
\protect\varphi ^{2}\rangle _{\text{t}}/m^{D-1}$, for $D=4$ as a function of
$mr$ for fixed value of $mL=1$ (left panel) and as a function of $\protect%
\beta $ for fixed values $mL=1$, $mr=0.25$ (right panel). The full and
dashed curves on the left panel correspond to untwisted and twisted fields,
respectively. The figures near the curves are the corresponding values of $q$%
.}
\label{fig1}
\end{figure}

Numerical examples below are given for the simplest 5-dimensional
Kaluza-Klein-type model ($D=4$) with a single extra dimension. In the left
panel of figure \ref{fig1} we depicted the topological part in the VEV of
the field squared as a function of $mr$ for $mL=1$ and for various values of
the parameter $q$ (figures near the curves). For $q=1$ the cosmic string is
absent and the VEV of the field squared is uniform. The full/dashed curves
correspond to untwsited/twisted fields ($\beta =0$ and $\beta =1/2$,
respectively). For a twisted field the complete VEV $\langle \varphi
^{2}\rangle $ is positive for points near the string and it is negative at
large distances. Hence, for some intermediate value of the radial coordinate
it vanishes. As it is seen, the presence of the cosmic string enhances the
vacuum polarization effects induced by the compactification of spatial
dimensions. In the right panel of figure \ref{fig1}, we plot the topological
part in the VEV of the field squared versus the parameter $\beta $ for $mL=1$
and $mr=0.25$. As before, the figures near the curves correspond to the
values of the parameter $q$. As it has been already mentioned, the
topological part is symmetric with respect to $\beta =1/2$. Note that the
intersection point of the graphs for different $q$ depends on the values of $%
mL$ and $mr$. For example, in the case $mL=0.25$ and $mr=0.25$ at the
intersection point we have $\beta \approx 0.19$ and $\langle \varphi
^{2}\rangle _{\text{t}}/m^{3}\approx 0.39$.

\section{Energy-momentum tensor}

\label{Sec:EMT}

Another important characteristic of the vacuum state is the VEV of the
energy-momentum tensor. For the evaluation of this quantity we use the
formula \cite{Saha04}
\begin{equation}
\langle T_{ik}\rangle =\lim_{x^{\prime }\rightarrow x}\partial _{i^{\prime
}}\partial _{k}W(x,x^{\prime })+\left[ \left( \xi -{1}/{4}\right)
g_{ik}\nabla _{l}\nabla ^{l}-\xi \nabla _{i}\nabla _{k}-\xi R_{ik}\right]
\langle \varphi ^{2}\rangle \ ,  \label{mvevEMT}
\end{equation}%
where for the spacetime under consideration the Ricci tensor,
$R_{ik}$, vanishes for points outside the string. The expression
for the energy-momentum tensor in (\ref{mvevEMT}) differs from the
standard one, given, for example, in \cite{Birr82}, by the term
which vanishes on the mass shell. By taking into account the
expressions for the Wightman function and the VEV of the field
squared, it can be seen that the vacuum energy-momentum tensor is
diagonal. Moreover, similar to the field squared, it is presented
in the decomposed form
\begin{equation}
\langle T_{i}^{j}\rangle =\langle T_{i}^{j}\rangle _{\text{s}}+\langle
T_{i}^{j}\rangle _{\text{t}},  \label{EMTDec}
\end{equation}%
where $\langle T_{i}^{j}\rangle _{\text{s}}$ is the corresponding VEV in the
geometry of a string without compactification and the part $\langle
T_{i}^{j}\rangle _{\text{t}}$ is induced by the nontrivial topology of the $z
$-direction. The topological part is finite and the renormalization is
reduced to that for the pure string part.

The topological part in the VEV of the energy-momentum tensor is found from (%
\ref{mvevEMT}), by making use of the expressions for the corresponding parts
in the Wightman function and the VEV of the field squared. After long but
straightforward calculations, for the topological part one finds
\begin{equation}
\langle T_{i}^{j}\rangle _{\text{t}}=\frac{4m^{D+1}}{(2\pi )^{(D+1)/2}}%
\sum_{l=1}^{\infty }\cos (2\pi l\beta )\left[ \sideset{}{'}{\sum}%
_{k=0}^{[q/2]}F_{i,l}^{j}(2mr,s_{k})-\frac{q\sin (q\pi )}{\pi }%
\int_{0}^{\infty }dy\frac{F_{i,l}^{j}(2mr,\cosh (y))}{\cosh (2qy)-\cos (q\pi
)}\right] ,  \label{EMTt}
\end{equation}%
where the functions for separate components are given by the expressions%
\begin{eqnarray}
F_{0,l}^{0}(u,v) &=&\left( 1-4\xi \right) u^{2}v^{4}f_{(D+3)/2}(w)-\left[
2\left( 1-4\xi \right) v^{2}+1\right] f_{(D+1)/2}(w),  \notag \\
F_{1,l}^{1}(u,v) &=&\left( 4\xi v^{2}-1\right) f_{(D+1)/2}(w),  \notag \\
F_{2,l}^{2}(u,v) &=&\left( 1-4\xi v^{2}\right) \left[
u^{2}v^{2}f_{(D+3)/2}(w)-f_{(D+1)/2}(w)\right] ,  \notag \\
F_{3,l}^{3}(u,v) &=&F_{0,l}^{0}(u,v)+\left( mlL\right) ^{2}f_{(D+3)/2}(w),
\label{Fij}
\end{eqnarray}%
with the function $f_{\nu }(x)$ defined by Eq. (\ref{fnu}), and we use the
notation
\begin{equation}
w=\sqrt{u^{2}v^{2}+\left( lmL\right) ^{2}}.  \label{yl}
\end{equation}%
For the components with $i>3$ one has (no summation) $%
F_{i}^{i}(u,v)=F_{0}^{0}(u,v)$. This relation is a direct consequence of the
invariance of the problem with respect to the boosts along the directions $%
x^{i}$, $i=4,...,D$. The topological part is symmetric with respect to $%
\beta =1/2$. In the presence of a constant gauge field the expression for
the VEV of the energy-momentum tensor is obtained from (\ref{EMTt}) by the
replacement $\beta \rightarrow \beta ^{\prime }$ with $\beta ^{\prime }$
given by (\ref{beta}). The topological part is a periodic function of the
component of the gauge field along the compact dimension.

In order to compare the contributions of the separate terms in (\ref{EMTDec}%
), here we give also the expression for the pure string part:
\begin{equation}
\langle T_{i}^{j}\rangle _{\text{s}}=\frac{2m^{D+1}}{(2\pi )^{(D+1)/2}}\left[
\sum_{k=1}^{[q/2]}F_{i,0}^{j}(2mr,s_{k})-\frac{q\sin (q\pi )}{\pi }%
\int_{0}^{\infty }dy\frac{F_{i,0}^{j}(2mr,\cosh (y))}{\cosh (2qy)-\cos (q\pi
)}\right] ,  \label{EMTs}
\end{equation}%
where the functions $F_{i,0}^{j}(2mr,s_{k})$ are given by (\ref{Fij}) with $%
l=0$. For integer values $q$, formula (\ref{EMTs}) is reduced to the one
given in \cite{Beze06}. The VEVs corresponding to (\ref{EMTs}) diverge on
the string as $1/r^{D+1}$. A procedure to cure this divergence is to
consider the string as having a nontrivial inner structure. In fact, in a
realistic point of view, the string has a characteristic core radius
determined by the energy scale where the symmetry of the system is
spontaneously broken.

\begin{figure}[tbph]
\begin{center}
\epsfig{figure=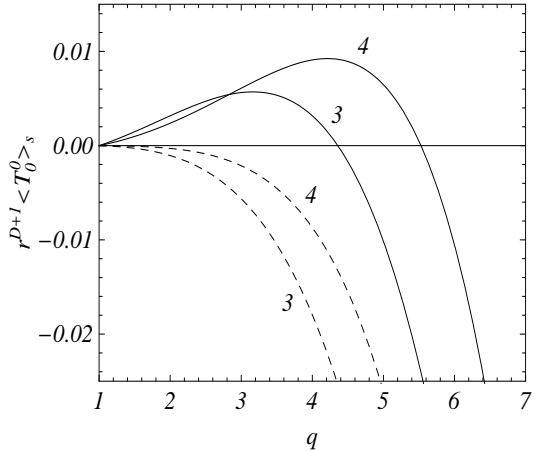, width=7.cm, height=6.cm}
\end{center}
\caption{The VEV of the energy density in the uncompactified string
geometry, $r^{D+1}\langle T_{0}^{0}\rangle _{\text{s}}$, as a function of $q$
for minimally (full curves) and conformally (dashed curves) coupled massless
scalar fields. The figures near the curves correspond to the values of $D$.}
\label{fig2}
\end{figure}

Various special cases of formula (\ref{EMTs}) can be found in the
literature. In particular, for a massless scalar field from (\ref{EMTs}) we
find the expression below:%
\begin{eqnarray}
\langle T_{i}^{j}\rangle _{\text{s}} &=&\frac{\Gamma ((D+1)/2)}{(4\pi
)^{(D+1)/2}r^{D+1}}\left\{ \left[ \frac{D-1}{D}g_{D}(q)-g_{D+2}(q)\right]
\text{diag}(1,1,-D,1,\ldots ,1)\right.  \notag \\
&&\left. -4(D-1)\left( \xi -\xi _{D}\right) g_{D}(q)\text{diag}(1,\frac{-1}{%
D-1},\frac{D}{D-1},1,\ldots ,1)\right\} ,  \label{EMTsm0}
\end{eqnarray}%
where the function $g_{D}(q)$ is defined by (\ref{gD}). In accordance with
the asymptotic estimate (\ref{gDlargeq}), for large values of $q$ the
expression on the right hand side of (\ref{EMTsm0}) is dominated by the term
with $g_{D+2}(q)$. The leading term does not depend on the curvature
coupling parameter and the corresponding energy density is always negative.
The energy density, $\langle T_{0}^{0}\rangle _{\text{s}}$, for a massless
field in arbitrary number of dimensions has been discussed previously in
\cite{Dowk87}. In the special case $D=3$, the expression (\ref{EMTsm0})
reduces to the one given in \cite{Frol87} (see also \cite{Hell86} for the
case of conformal coupling and \cite{Guim94} for a string which carries an
internal magnetic flux). In this case and for a conformally coupled field
the corresponding energy density is always negative. For a minimally coupled
field the energy density is positive for $q^{2}<19$ and it is negative for $%
q^{2}>19$. In figure \ref{fig2} we plot the energy density corresponding to (%
\ref{EMTsm0}) as a function of $q$ for $D=3,4$ (figures near the curves) for
minimally (full curves) and conformally (dashed curves) coupled massless
fields. In the discussion below we shall be mainly concerned with the
topological part.

It can be easily checked that the topological part satisfies the covariant
conservation equation for the energy-momentum tensor: $\nabla _{j}\langle
T_{i}^{j}\rangle _{\text{t}}=0$. In the geometry under consideration, the
latter is reduced to the equation $\langle T_{2}^{2}\rangle _{\text{t}%
}=\partial _{r}\left( r\langle T_{1}^{1}\rangle _{\text{t}}\right) $. In
addition, the topological part obeys the trace relation%
\begin{equation}
\langle T_{i}^{i}\rangle _{\text{t}}=D(\xi -\xi _{D})\nabla _{i}\nabla
^{i}\langle \varphi ^{2}\rangle _{\text{t}}+m^{2}\langle \varphi ^{2}\rangle
_{\text{t}}.  \label{trace}
\end{equation}%
In particular, it is traceless for a conformally coupled massless field.

Let us consider special cases of the general formula (\ref{EMTt}). For
integer values of $q$ one finds:%
\begin{equation}
\langle T_{i}^{j}\rangle _{\text{t}}=\frac{2m^{D+1}}{(2\pi )^{(D+1)/2}}%
\sum_{k=0}^{q-1}\sum_{l=1}^{\infty }\cos (2\pi l\beta
)F_{i,l}^{j}(2mr,s_{k}).  \label{EMTtsp}
\end{equation}%
In the case of a massless field and general values of $q$ the formula (\ref%
{EMTt}) reduces to%
\begin{eqnarray}
\langle T_{i}^{j}\rangle _{\text{t}} &=&\frac{2\Gamma ((D+1)/2)}{\pi
^{(D+1)/2}L^{D+1}}\sum_{l=1}^{\infty }\frac{\cos (2\pi l\beta )}{l^{D+1}}%
\bigg[\sideset{}{'}{\sum}_{k=0}^{[q/2]}\frac{F_{i}^{(0)j}(2r/lL,s_{k})}{%
[(2rs_{k}/lL)^{2}+1]^{(D+3)/2}}  \notag \\
&&-\frac{q\sin (q\pi )}{\pi }\int_{0}^{\infty }dy\frac{F_{i}^{(0)j}(2r/lL,%
\cosh (y))}{\cosh (2qy)-\cos (q\pi )}[(2r/lL)^{2}\cosh ^{2}(y)+1]^{-(D+3)/2}%
\bigg],  \label{EMTtm0}
\end{eqnarray}%
with the notations%
\begin{eqnarray}
F_{0}^{(0)0}(u,v) &=&\left( 1-4\xi \right) v^{2}\left[ (D-1)v^{2}u^{2}-2%
\right] -u^{2}v^{2}-1,  \notag \\
F_{1}^{(0)1}(u,v) &=&\left( 4\xi v^{2}-1\right) \left( u^{2}v^{2}+1\right) ,
\notag \\
F_{2}^{(0)2}(u,v) &=&\left( 1-4\xi v^{2}\right) \left( Du^{2}v^{2}-1\right) ,
\notag \\
F_{3}^{(0)3}(u,v) &=&F_{0}^{(0)0}(u,v)+D+1,  \label{Fijl}
\end{eqnarray}%
and (no summation) $F_{i}^{(0)i}(u,v)=F_{0}^{(0)0}(u,v)$ for $i>3$.

\begin{figure}[tbph]
\begin{center}
\epsfig{figure=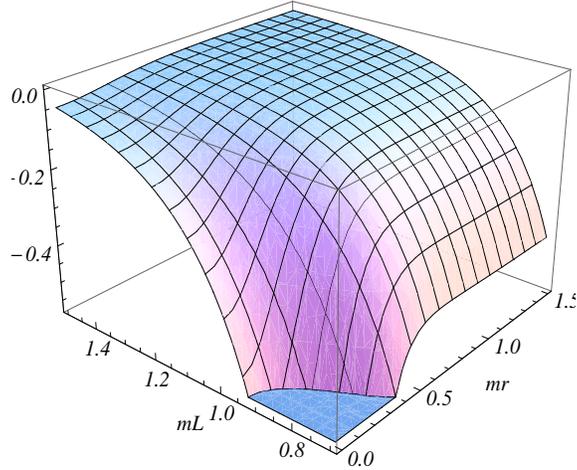, width=7.5cm, height=6.5cm}
\end{center}
\caption{The topological part in the VEV of the energy density, $\langle
T_{0}^{0}\rangle _{\text{t}}/m^{D+1}$, for a minimally coupled untwisted
scalar field in 5-dimensional cosmic string spacetime with $q=3$, as a
function of $mr$ and $mL$.}
\label{fig3}
\end{figure}

Now we consider the asymptotics for the VEV of the energy-momentum tensor.
When $r\gg L$, the dominant contribution in (\ref{EMTt}) comes from the term
with $k=0$:%
\begin{eqnarray}
\langle T_{i}^{j}\rangle _{\text{t}} &\approx &\langle T_{i}^{j}\rangle _{%
\text{t}}^{\text{(M)}}=-\frac{2m^{D+1}}{(2\pi )^{(D+1)/2}}\sum_{l=1}^{\infty
}\cos (2\pi l\beta )f_{(D+1)/2}(lmL)  \notag \\
&&\times \text{diag}(1,1,1,-D+\frac{f_{(D-1)/2}(lmL)}{f_{(D+1)/2}(lmL)}%
,1,\ldots ,1),  \label{EMTtlr}
\end{eqnarray}%
where $\langle T_{i}^{j}\rangle _{\text{t}}^{\text{(M)}}$ is the VEV in the
Minkowski spacetime with a compact dimension of the length $L$. Note that
the latter does not depend on the curvature coupling parameter. From (\ref%
{EMTtlr}) it follows that at large distances from the string the topological
part in the energy density is negative/positive for untwisted/twisted scalar
fields. At large distances the topological part dominates and the same is
the case for the total VEV. On the string we have
\begin{eqnarray}
\langle T_{i}^{j}\rangle _{\text{t},r=0} &=&q\langle T_{i}^{j}\rangle _{%
\text{t}}^{\text{(M)}}+\frac{4m^{D+1}F_{i}^{j}}{(2\pi )^{(D+1)/2}}%
\sum_{l=1}^{\infty }\cos (2\pi l\beta )f_{(D+1)/2}(lmL)  \notag \\
&&\times \left[ \sideset{}{'}{\sum}_{k=0}^{[q/2]}s_{k}^{2}-\frac{q\sin (q\pi
)}{\pi }\int_{0}^{\infty }dy\frac{\cosh ^{2}(y)}{\cosh (2qy)-\cos (q\pi )}%
\right] ,  \label{EMTtaxis}
\end{eqnarray}%
where the notations are as follows (no summation):%
\begin{equation}
F_{i}^{i}=2\left( 4\xi -1\right) ,\;i=0,3,4,\ldots
,D,\;F_{1}^{1}=F_{2}^{2}=4\xi .  \label{Fijaxis}
\end{equation}%
For both conformally and minimally coupled fields the energy density
corresponding to (\ref{EMTtaxis}) is negative/positive for untwisted/twisted
fields. For integer values of $q$, the expression in the square brackets in (%
\ref{EMTtaxis}) is equal to $q/4$. The pure string part of the VEV diverges
on the string as $1/r^{D+1}$ and, hence, it dominates for points near the
string, $r\ll L$. Combining these features, we see that for a minimally
coupled untwisted scalar field the vacuum energy is negative at large
distances from the string and it is positive near the string for $q^{2}<19$
and $D\geqslant 3$. In figure \ref{fig3} we plot the topological part in the
vacuum energy density, $\langle T_{0}^{0}\rangle _{\text{t}}/m^{D+1}$, as a
function of the distance from the string and of the length of the compact
dimension, for an untwisted scalar field ($\beta =0$) in $D=4$ cosmic string
spacetime with $q=3$. For a twisted scalar field the corresponding energy
density is positive.

\begin{figure}[tbph]
\begin{tabular}{cc}
\epsfig{figure=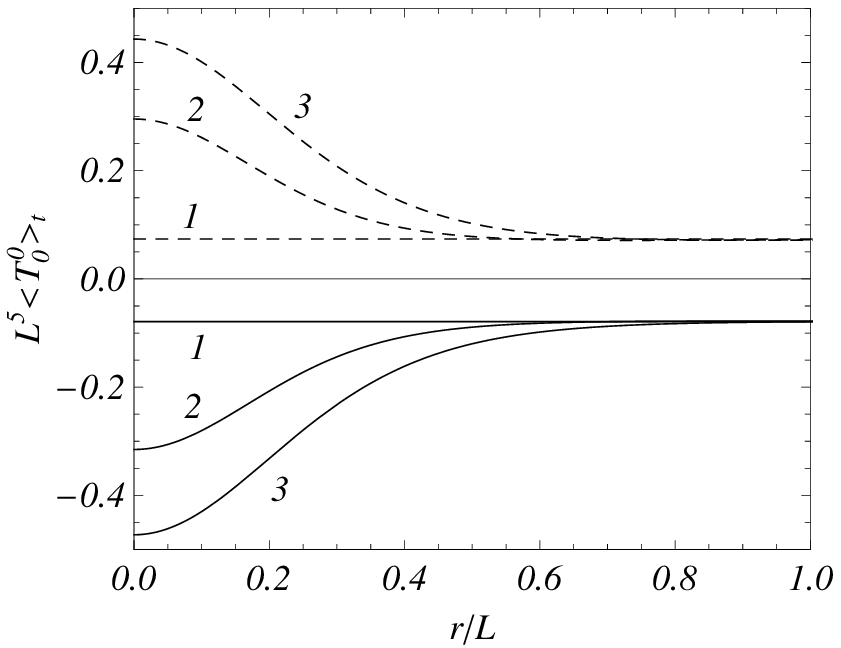, width=7.cm, height=6.cm} & \quad %
\epsfig{figure=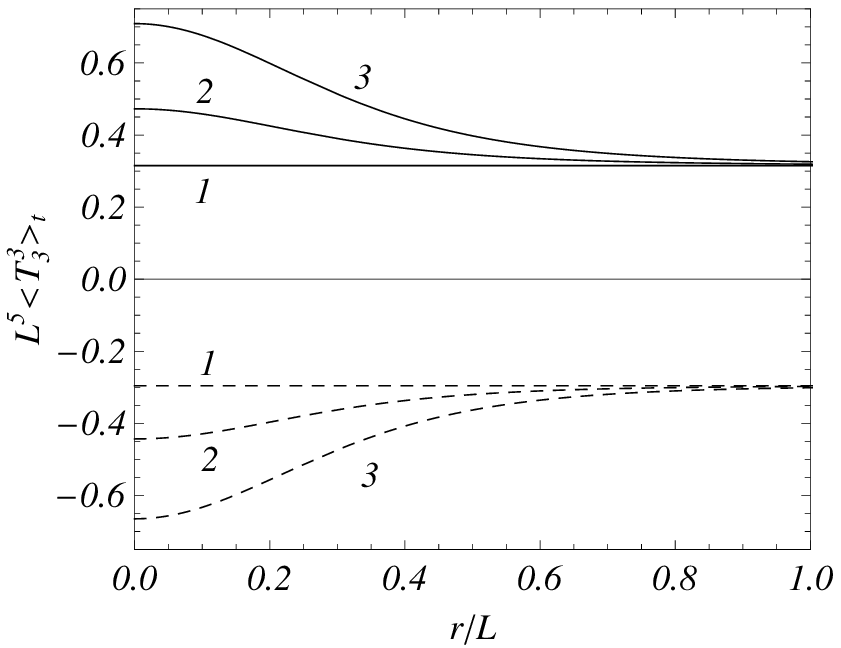, width=7.cm, height=6.cm}%
\end{tabular}%
\caption{The topological parts in the VEV of the energy density (left panel)
and the stress along the compact dimension, as functions of $r/L$ for a
minimally coupled massless scalar field. The full and dashed curves
correspond to untwisted and twisted fields, respectively. The figures near
the curves are the values of the parameter $q$.}
\label{fig4}
\end{figure}

For a massless field the quantity $L^{D+1}\langle T_{i}^{j}\rangle _{\text{t}%
}$ is a function of the ratio $r/L$. In figure \ref{fig4} we present the
corresponding energy density (left panel) and the stress along the compact
dimension (right panel) in the case of a $D=4$ minimally coupled scalar
field for various values of the parameter $q$ (figures near the curves). The
full/dashed curves correspond to untwsited/twisted scalar fields. In the
case $q=1$ the cosmic string is absent and the corresponding VEVs are
uniform. The graphs for a conformally coupled field are similar to those
given in figure \ref{fig4}. Note that for the topological part in the vacuum
effective pressure along the $j$-th direction we have (no summation) $p_{%
\text{t},j}=-$ $\langle T_{j}^{j}\rangle _{\text{t}}$, and, hence, for the
example corresponding to figure \ref{fig4} both the energy density and the
pressure along the compact dimension are negative/positive for
untwisted/twisted fields.

\begin{figure}[tbph]
\begin{center}
\epsfig{figure=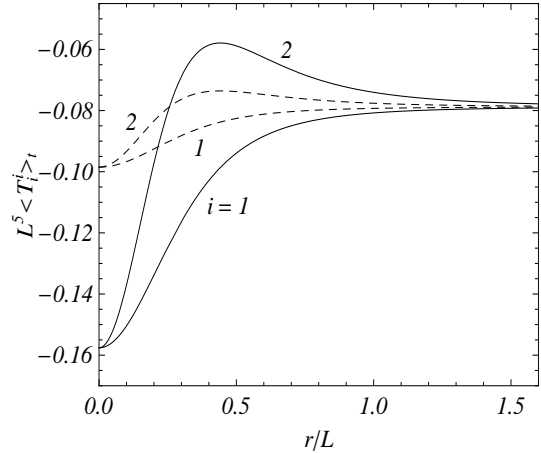, width=7.cm, height=6.cm}
\end{center}
\caption{The topological parts in the radial and azimuthal stresses for a $%
D=4$ minimally coupled massless field as a function of the ratio $r/L $ in
the geometry of a $D=4$ cosmic string with $q=2$.}
\label{fig5}
\end{figure}

The topological parts in the radial and azimuthal stresses are plotted in
figure \ref{fig5} versus $r/L$ for minimally (full curves) and conformally
(dashed curves) coupled untwisted fields in the geometry of a $D=4$ cosmic
string with $q=2$. Note that the azimuthal stress is not a monotonic
function in both cases. As it is seen, the corresponding effective pressures
are positive. For a twisted field the graphs have similar structure with the
signs changed.

\begin{figure}[tbph]
\begin{center}
\epsfig{figure=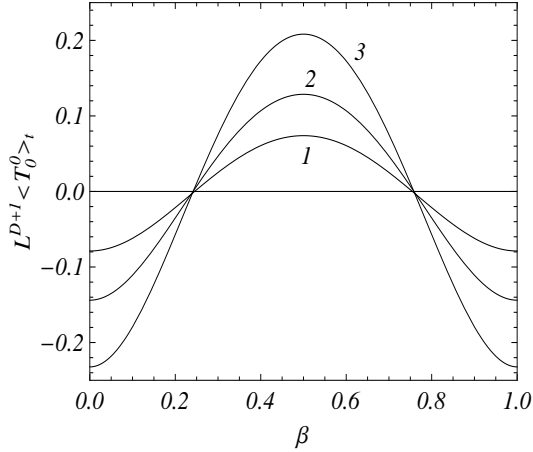, width=7.cm, height=6.cm}
\end{center}
\caption{The topological part in the VEV of the energy density for a $D=4$
minimally coupled massless field as a function of the phase parameter $%
\protect\beta $ for $r/L=0.3$. The figures near the curves correspond to the
values of $q$.}
\label{fig6}
\end{figure}

The dependence of the energy density on the parameter $\beta $ in the
quasiperiodicity condition along the compact dimension is presented in
figure \ref{fig6} for a minimally coupled massless scalar field in $D=4$.
The graphs are plotted for $r/L=0.3$ and for various values of $q$ (numbers
near the curves). They are symmetric with respect to $\beta =1/2$.

\section{Vacuum energy}

\label{Sec:VacEn}

As we have seen before, the energy density corresponding to the pure string
part diverges on the string as $1/r^{D+1}$. Consequently, the corresponding
total vacuum energy is divergent. We can evaluate the total vacuum energy in
the region $r\geqslant r_{0}>0$ per unit volume in the subspace $%
(x^{3},x^{4},\ldots ,x^{D})$, defined as $E_{0,r\geqslant r_{0}}^{\text{(s)}%
}=\phi _{0}\int_{r_{0}}^{\infty }dr\,r\langle T_{0}^{0}\rangle _{\text{s}}$.
By using the recurrence relation for the modified Bessel function and the
formula (see \cite{Prud86}) $\int_{a}^{\infty }dx\,xf_{\nu }(x)=f_{\nu
-1}(a) $, with the function $f_{\nu }(x)$ defined in (\ref{fnu}), one finds%
\begin{equation}
E_{0,r\geqslant r_{0}}^{\text{(s)}}=\frac{m^{D-1}\phi _{0}}{2(2\pi
)^{(D+1)/2}}\left[ \sum_{k=1}^{[q/2]}F(2mr_{0}s_{k},s_{k})-\frac{q\sin (q\pi
)}{\pi }\int_{0}^{\infty }dy\frac{F(2mr_{0}\cosh (y),\cosh (y))}{\cosh
(2qy)-\cos (q\pi )}\right] ,  \label{Es0}
\end{equation}%
with the notation%
\begin{equation}
F(u,v)=\left( 1-4\xi \right) u^{2}f_{(D+1)/2}(u)-f_{(D-1)/2}(u)/v^{2}.
\label{Fuv}
\end{equation}%
For a massless field this formula is reduced to%
\begin{equation}
E_{0,r\geqslant r_{0}}^{\text{(s)}}=\frac{\Gamma ((D-1)/2)}{4(4\pi
)^{(D-1)/2}qr_{0}^{D-1}}\left[ \left( 1-4\xi \right) (D-1)g_{D}(q)-g_{D+2}(q)%
\right] .  \label{Es0m0}
\end{equation}%
Of course, this result could also be obtained directly from (\ref{EMTsm0}).
For $mr_{0}\gg 1$ and for fixed value of $q$, the dominant contribution to (%
\ref{Es0}) comes from the first term in the right hand side of Eq. (\ref{Fuv}%
) and the vacuum energy is positive for $\xi <1/4$. In particular, this is
the case for both minimally and conformally coupled fields. In the case $%
mr_{0}\ll 1$, the leading term in the asymptotic expansion of the vacuum
energy is given by (\ref{Es0m0}). For large values of $q$ the second term in
the square brackets of (\ref{Es0m0}) dominates and the vacuum energy is
negative with independence of the curvature coupling parameter. For $%
q\gtrsim 1$, the vacuum energy given by (\ref{Es0m0}) remains negative for a
conformally coupled field and becomes positive for a minimally coupled field.

Now we turn to the topological part of the vacuum energy. The corresponding
energy-momentum tensor can be further decomposed as
\begin{equation}
\langle T_{i}^{j}\rangle _{\text{t}}=\langle T_{i}^{j}\rangle _{\text{t}}^{%
\text{(M)}}+\langle T_{i}^{j}\rangle _{\text{t}}^{\text{(s)}},
\label{DecTop}
\end{equation}%
where the second term on the right hand side is the correction due to the
presence of the string. The expression for $\langle T_{i}^{j}\rangle _{\text{%
t}}^{\text{(s)}}$ is obtained from (\ref{EMTt}) by subtracting the part
corresponding to the term $k=0$ (the latter coincides with $\langle
T_{i}^{j}\rangle _{\text{t}}^{\text{(M)}}$). For the correction in the
topological part of the vacuum energy per unit volume in the subspace $%
(x^{4},\ldots ,x^{D})$, induced by the string, we have
\begin{equation}
E_{\text{t}}^{\text{(s)}}=\int_{0}^{\infty }dr\,r\int_{0}^{\phi _{0}}d\phi
\int_{0}^{L}dz\,\langle T_{0}^{0}\rangle _{\text{t}}^{\text{(s)}}.
\label{EST}
\end{equation}%
By taking into account that $\langle T_{0}^{0}\rangle _{\text{t}}^{\text{(s)}%
}$ is given by the expression (\ref{EMTt}) omitting the $k=0$ term, the
integral over $r$ is evaluated by using the formula \cite{Prud86}%
\begin{equation}
\int_{a}^{\infty }dx\,x(x^{2}-a^{2})^{\beta -1}f_{\nu }(cx)=2^{\beta
-1}c^{-2\beta }f_{\nu -\beta }(ac),  \label{IntForm2}
\end{equation}%
with the function $f_{\nu }(x)$ defined in Eq. (\ref{fnu}). As a result, the
dependence on the parameter $q$ is factorized in the form of $g_{3}(q)/q$,
where the function $g_{D}(q)$ is given by (\ref{gD}). By taking into account
the expression (\ref{g35}) for $g_{3}(q)$, we find the final expression for
the vacuum energy:%
\begin{equation}
E_{\text{t}}^{\text{(s)}}=-\frac{q^{2}-1}{6q}\frac{m^{D-1}L}{(2\pi
)^{(D-1)/2}}\sum_{l=1}^{\infty }\cos (2\pi l\beta )f_{(D-1)/2}(lmL).
\label{EST2}
\end{equation}%
As it is seen, the total energy does not depend on the curvature coupling
parameter. It is negative/positive for untwisted/twisted scalar fields.

\begin{figure}[tbph]
\begin{center}
\epsfig{figure=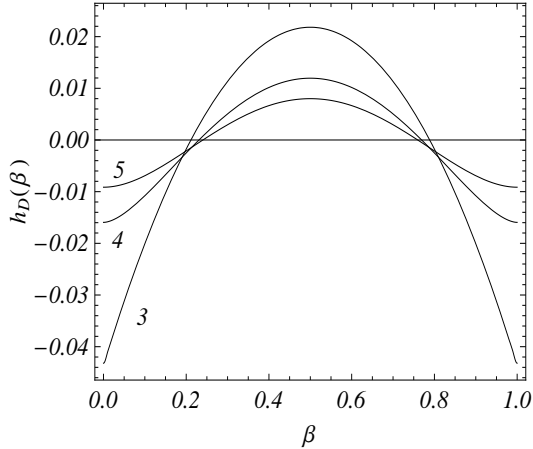, width=7.cm, height=6.cm}
\end{center}
\caption{The function $h_{D}(\protect\beta )$ in (\protect\ref{ESTm0}) for $%
D=3,4,5$ (figures near the curves).}
\label{fig7}
\end{figure}

For a massless field we find%
\begin{equation}
E_{\text{t}}^{\text{(s)}}=\frac{q^{2}-1}{qL^{D-2}}h_{D}(\beta ),
\label{ESTm0}
\end{equation}%
where%
\begin{equation}
h_{D}(\beta )=-\frac{\Gamma ((D-1)/2)}{12\pi ^{(D-1)/2}}\sum_{l=1}^{\infty }%
\frac{\cos (2\pi l\beta )}{l^{D-1}}.  \label{hD}
\end{equation}%
For a massive field, the expression (\ref{ESTm0}) gives the leading term in
the asymptotic expansion for $mL\ll 1$. For odd values of $D$, the series in
(\ref{hD}) is given in terms of the Bernoulli polynomials:
\begin{equation}
h_{D}(\beta )=\frac{(-1)^{(D-1)/2}\pi ^{D/2}}{12(D-1)\Gamma (D/2)}%
B_{D-1}(\beta ).  \label{ESTm0odd}
\end{equation}%
In figure \ref{fig7} we plotted the function $h_{D}(\beta )$ for $D=3,4,5$
(figures near the curves). As in the case of the vacuum densities, this
function is symmetric with respect to $\beta =1/2$.

\section{Conclusion}

\label{sec:Conc}

In the present paper we have investigated the one-loop quantum effects for a
massive scalar field with general curvature coupling parameter, induced by
the compactification of spatial dimensions in a generalized cosmic string
spacetime. It is assumed that along the compact dimension the field obeys
quasiperiodicity condition with an arbitrary phase. As the first step for
the investigation of vacuum densities we evaluate the positive frequency
Wightman function. This function gives comprehensive insight into vacuum
fluctuations and determines the response of a particle detector of the
Unruh-DeWitt type in a given state of motion. For a massive field and for
general value of the planar angle deficit, the Wightman function is given by
formula (\ref{WF4}). The $l=0$ term in this expression corresponds to the
Wightman function in the geometry of a cosmic string without
compactification and, hence, the topological part is explicitly extracted.
As the compactification under consideration does not change the local
geometry, in this way the renormalization for the VEVs in the coincidence
limit is reduced to that for the standard cosmic string geometry without
compactification. For integer values of the parameter $q$, the Wightman
function is expressed as an image sum of the corresponding function in the
Minkowski spacetime with a compact dimension and is given by Eq. (\ref{WF4Sp}%
).

The VEV of the field squared is decomposed as the sum of the pure
string part and the correction due to the compactification. For a
massive field the string part is given by (\ref{phi2s}). Since the
geometry is locally flat, this part does not depend on the
curvature coupling parameter. It is positive for $q>1$ and
diverges on the string like $1/r^{D-1}$. This divergence may be
regularized considering a more realistic model of the string with
nontrivial inner structure. The topological part in the VEV\ of
the field squared is given by the expression (\ref{phi2t}) and it
is finite everywhere including the points on the string. In
dependence of the value of the phase $\beta $ in the
quasiperiodicity condition, this part can be either positive or
negative. In particular, the topological part is positive for an
untwisted scalar and it is negative for a twisted scalar. At
distances from the string larger than the length of the compact
dimension, the topological part in the VEV of the field squared
approaches to the corresponding quantity in the Minkowski
spacetime with a compact dimension. For points near the string we
have the simple asymptotic relation $\langle \varphi
^{2}\rangle _{\text{t}}\approx q\langle \varphi ^{2}\rangle _{\text{t}}^{%
\text{(M)}}$.

The VEV of the energy-momentum tensor is investigated in section \ref%
{Sec:EMT}. This VEV is diagonal and, similar to the case of the field
squared, it is decomposed into the pure string and topological parts, given
by expressions (\ref{EMTs}) and (\ref{EMTt}), respectively. We have
explicitly checked that the topological part satisfies the covariant
conservation equation and its trace is related to the VEV of the field
squared by the standard formula. For a massive field and for general value
of the parameter $q$, we give a closed expression for the pure string part
in the VEV of the energy-momentum tensor in arbitrary number of dimensions.
The latter generalizes various special cases previously discussed in the
literature. At large distances from the string, the topological part
coincides with the corresponding result in the Minkowski spacetime and it
dominates in the total VEV. In this limit the VEV of the energy-momentum
tensor does not depend on the curvature coupling parameter and the
corresponding energy density is negative/positive for untwisted/twisted
scalar fields. The topological part is finite on the string and for points
near the string the leading term in the corresponding asymptotic expansion
is given by (\ref{EMTtaxis}). The pure string part in the VEV of the energy
density diverges on the string as $1/r^{D+1}$ and near the string it
dominates in the total VEV. For a conformally coupled scalar field the
corresponding energy density is negative, whereas for a minimally coupled
field the energy density is positive for small values of the parameter $q$
and becomes negative for large values of $q$. The numerical examples are
given for the simplest Kaluza-Klein-type model with a single extra
dimension. They show that he nontrivial topology due to the cosmic string
enhances the vacuum polarization effects induced by the compactness of
spatial dimensions for both the field squared and the vacuum energy density.
For a charged scalar field, in the presence of a constant gauge field the
expression for the topological parts are obtained from the formulas given
above by the replacement $\beta \rightarrow \beta ^{\prime }$ with $\beta
^{\prime }$ defined by (\ref{beta}). In this case the topological parts are
periodic functions of the component of the gauge field along the compact
dimension. This is an analog of the Aharonov-Bohm effect.

As a result of the non-integrable divergence of the energy density in the
pure string part on the string, the corresponding total vacuum energy is
divergent. In section \ \ref{Sec:VacEn} we give a closed expression, Eq. (%
\ref{Es0}), for the vacuum energy in the region $r\geqslant r_{0}>0$. The
topological part in the VEV of the energy-momentum tensor can be further
decomposed into the Minkowskian and string induced parts. The latter is
finite on the string and vanishes at large distances from the string. As a
result, the total vacuum energy corresponding to this part is finite. This
energy is given by the expressions (\ref{EST2}) and (\ref{ESTm0}) for
massive and massless fields, respectively. In these expressions the
dependence on the parameter $q$ is simply factorized. The string induced
part in the topological energy does not depend on the curvature coupling
parameter and it is negative/positive for untwisted/twisted scalars.

In a way similar to that described above, one can consider the topological
Casimir effect for a cosmic string in de Sitter spacetime with compact
spatial dimensions. The vacuum polarization effects induced by the presence
of the string in uncompactified de Sitter spacetime have been recently
discussed in Ref. \cite{Beze09}. The topological Casimir densities in de
Sitter spacetime with toroidally compactified spatial dimensions are
investigated in Ref. \cite{Saha08}. It has been shown that the curvature of
the background spacetime decisively influences the behavior of the
topological parts in the VEVs of the field squared and the energy density
for lengths of compact dimensions larger than the curvature scale of the
spacetime.

In this paper the string geometry is taken as a static, given
classical background for quantum matter fields. This approach
follows the main part of the papers where the influence of the
string on quantum matter is investigated (see references
\cite{Hell86}-\cite{Guim94}). Of course, in a more complete
approach the dynamics of the cosmic string should be taken into
account. In the simplest model, the cosmic string dynamics can be
described by the Nambu action (see, for instance, \cite{Vile94}).
If the scalar field under consideration interacts with the Higgs
field inside the string core, then, within this model, the total
action will contain also the term describing the interaction of
the scalar field with the vibrational modes of the string. This
would be the further development of the model under discussion.
The results obtained in the present paper are the first step to
this more general problem. Another development would be the
investigation of the back-reaction effects of the quantum
energy-momentum tensor on the gravitational field of the cosmic
string. For the geometry of infinitely thin straight cosmic string
the back-reaction for conformal fields has been discussed in
\cite{Iell97,Hisc87} by using the linearized semiclassical
Einstein equations. It would also be interesting to generalize the
vacuum polarization calculations of the present paper for the
models with nontrivial string core. For a general cylindrically
symmetric static model of the string core with finite support this
can be done in a way similar to that used in \cite{Beze06} for the
geometry of a straight cosmic string.

\section*{Acknowledgments}

AAS was supported by CAPES Program. ERBM thanks Conselho Nacional de
Desenvolvimento Cient\'{\i}fico e Tecnol\'{o}gico (CNPq) for partial
financial support.


\begin{thebibliography}{99}
\bibitem{Kibb80} T.W.B. Kibble, Phys. Rep. \textbf{67}, 183 (1980); A.
Vilenkin, Phys. Rep. \textbf{121}, 263 (1985).

\bibitem{Vile94} A. Vilenkin and E.P.S. Shellard, \textit{Cosmic Strings and
Other Topological Defects} (Cambridge University Press, Cambridge, England,
1994).

\bibitem{Damo00} T. Damour and A. Vilenkin, Phys. Rev. Lett. \textbf{85},
3761 (2000).

\bibitem{Bhat00} P. Bhattacharjee and G. Sigl, Phys. Rept. \textbf{327}, 109
(2000).

\bibitem{Bere01} V. Berezinsky, B. Hnatyk, and A. Vilenkin, Phys. Rev. D
\textbf{64}, 043004 (2001).

\bibitem{Sara02} S. Sarangi and S.H.H. Tye, Phys. Lett. B \textbf{536}, 185
(2002); E.J. Copeland, R.C. Myers, and J. Polchinski, JHEP \textbf{0406},
013 (2004); G. Dvali and A. Vilenkin, JCAP \textbf{03}, 010 (2004).

\bibitem{Hell86} T.M. Helliwell and D.A. Konkowski, Phys. Rev. D \textbf{34}%
, 1918 (1986).

\bibitem{Line87} B. Linet, Phys. Rev. D \textbf{35}, 536 (1987).

\bibitem{Frol87} V.P. Frolov and E.M. Serebriany, Phys. Rev. D \textbf{35},
3779 (1987).

\bibitem{Dowk87} J.S. Dowker, Phys. Rev. D \textbf{36}, 3095 (1987); J.S.
Dowker, Phys. Rev. D \textbf{36}, 3742 (1987).

\bibitem{Davi88} P.C.W. Davies and V. Sahni, Class. Quantum Grav. \textbf{5}%
, 1 (1988).

\bibitem{Smit89} A.G. Smith, in \textit{The Formation and Evolution of
Cosmic Strings}, Proceedings of the Cambridge Workshop, Cambridge, England,
1989, edited by G.W. Gibbons, S.W. Hawking, and T. Vachaspati (Cambridge
University Press, Cambridge, England, 1990).

\bibitem{Mats90} G.E.A. Matsas, Phys. Rev. D \textbf{41}, 3846 (1990).

\bibitem{Alle90} B. Allen and A.C. Ottewill, Phys. Rev. D \textbf{42}, 2669
(1990); B. Allen, J.G. Mc Laughlin, and A.C. Ottewill, Phys. Rev. D \textbf{%
45}, 4486 (1992); B. Allen, B.S. Kay, and A.C. Ottewill, Phys. Rev. D
\textbf{53}, 6829 (1996).

\bibitem{Sour92} T. Souradeep and V. Sahni, Phys. Rev. D \textbf{46}, 1616
(1992).

\bibitem{Shir92} K. Shiraishi and S. Hirenzaki, Class. Quantum Grav. \textbf{%
9}, 2277 (1992).

\bibitem{Beze94} V.B. Bezerra and E.R. Bezerra de Mello, Class. Quantum
Grav. \textbf{11}, 457 (1994); E.R. Bezerra de Mello, Class. Quantum Grav.
\textbf{11}, 1415 (1994).

\bibitem{Cogn94} G. Cognola, K. Kirsten, and L. Vanzo, Phys. Rev. D \textbf{%
49}, 1029 (1994).

\bibitem{More95} E.S. Moreira Jnr, Nucl. Phys. B \textbf{451}, 365 (1995).

\bibitem{Iell97} D. Iellici, Class. Quantum Grav. \textbf{14}, 3287 (1997).

\bibitem{Khus99} N.R. Khusnutdinov and M. Bordag, Phys. Rev. D \textbf{59},
064017 (1999).

\bibitem{BezeKh06} V.B. Bezerra and N.R. Khusnutdinov, Class. Quantum Grav.
\textbf{23}, 3449 (2006).

\bibitem{Guim94} M.E.X. Guimar\~{a}es and B. Linet, Commun. Math. Phys.
\textbf{165}, 297 (1994); M.E.X. Guimar\~{a}es, Class. Quantum Grav. \textbf{%
12}, 1705 (1995); L. Sriramkumar, Class. Quantum Grav. \textbf{18}, 1015
(2001); J. Spinelly and E.R. Bezerra de Mello, Class. Quantum Grav. \textbf{%
20}, 873 (2003); Yu.A. Sitenko and N.D. Vlasii, Class. Quantum Grav. \textbf{%
26}, 195009 (2009).

\bibitem{Cast09} A.H. Castro Neto, F. Guinea, N.M.R. Peres, K.S. Novoselov,
and A.K. Geim, Rev.Mod. Phys. \textbf{81}, 109 (2009).

\bibitem{Most97} V.M. Mostepanenko and N.N. Trunov, \textit{The Casimir
Effect and Its Applications} (Oxford University Press, Oxford, 1997).

\bibitem{Eliz94} E. Elizalde, S.D. Odintsov, A. Romeo, A.A. Bytsenko, and S.
Zerbini, \textit{Zeta Regularization Techniques with Applications} (World
Scientific, Singapore, 1994); K.A. Milton \textit{The Casimir Effect:
Physical Manifestation of Zero-Point Energy} (World Scientific, Singapore,
2002); M. Bordag, G. L. Klimchitskaya, U. Mohideen, and V.M. Mostepanenko,
\textit{Advances in the Casimir Effect} (Oxford University Press, Oxford,
2009).

\bibitem{Chen06} H.B. Cheng, Phys. Lett. B \textbf{643}, 311 (2006); H.B.
Cheng, Phys. Lett. B \textbf{668}, 72 (2008); S.A. Fulling and K. Kirsten,
Phys. Lett. B \textbf{671}, 179 (2009); K. Kirsten and S.A. Fulling, Phys.
Rev. D \textbf{79}, 065019 (2009); E. Elizalde, S.D. Odintsov, and A.A.
Saharian, Phys. Rev. D \textbf{79}, 065023 (2009); L.P. Teo, Phys. Lett. B
\textbf{672}, 190 (2009); L.P. Teo, Nucl. Phys. B \textbf{819}, 431 (2009);
L.P. Teo, JHEP \textbf{0906}, 076 (2009); L.P. Teo, JHEP \textbf{0911}, 095
(2009).

\bibitem{Popp04} K. Poppenhaeger, S. Hossenfelder, S. Hofmann, and M.
Bleicher, Phys. Lett. B \textbf{582}, 1 (2004); A. Edery and V.N.
Marachevsky, Phys. Rev. D \textbf{78}, 025021 (2008); A. Edery and V.N.
Marachevsky, JHEP \textbf{0812}, 035 (2008); F. Pascoal, L.F.A. Oliveira,
F.S.S. Rosa, and C. Farina, Braz. J. Phys. \textbf{38}, 581 (2008); L.
Perivolaropoulos, Phys. Rev. D \textbf{77}, 107301 (2008).

\bibitem{Bell09} S. Bellucci and A.A. Saharian, Phys. Rev. D \textbf{80},
105003 (2009); E. Elizalde, S.D. Odintsov, and A.A. Saharian, Phys. Rev. D
\textbf{83}, 105023 (2011).

\bibitem{Brev95} I. Brevik and T. Toverud, Class. Quantum Grav. \textbf{12},
1229 (1995).

\bibitem{Beze06} E.R. Bezerra de Mello, V.B. Bezerra, A.A. Saharian and A.S.
Tarloyan, Phys. Rev. D \textbf{74}, 025017 (2006).

\bibitem{Beze07} E.R. Bezerra de Mello, V.B. Bezerra, and A.A. Saharian,
Phys. Lett. B \textbf{645}, 245 (2007); E. R. Bezerra de Mello, V. B.
Bezerra, A. A. Saharian, and A. S. Tarloyan, Phys. Rev. D \textbf{78},
105007 (2008); G. Fucci and K. Kirsten, JHEP \textbf{1103}, 016 (2011); E.R.
Bezerra de Mello and A.A. Saharian, Class. Quantum Grav. \textbf{28}, 145008
(2011); G. Fucci and K. Kirsten, J. Phys. A \textbf{44}, 295403 (2011).

\bibitem{Volo98} G.E. Volovik, Pisma Zh. Eksp. Teor. Fiz. \textbf{67}, 666
(1998) [JETP Lett. \textbf{67}, 698 (1998)].

\bibitem{Kris97} A. Krishnan, et al, Nature \textbf{388}, 451 (1997); S.N.
Naess, A. Elgsaeter, G. Helgesen and K.D. Knudsen, Sci. Technol. Adv. Mater.
\textbf{10}, 065002 (2009).

\bibitem{Beze10} E.R. Bezerra de Mello, V.B. Bezerra, A.A. Saharian, and
V.M. Bardeghyan, Phys. Rev. D \textbf{82}, 085033 (2010); S. Bellucci, E.R.
Bezerra de Mello, and A.A. Saharian, Phys. Rev. D \textbf{83}, 085017 (2011).

\bibitem{Birr82} N.D. Birrell and P.C.W. Davies, \textit{Quantum Fields in
Curved Space} (Cambridge University Press, Cambridge, England, 1982).

\bibitem{Taga86} S. Tagaki, Prog. Theor. Phys. Suppl. \textbf{88}, 1 (1986).

\bibitem{Bell10} S. Bellucci, A.A. Saharian, and V.M. Bardeghyan, Phys. Rev.
D \textbf{82}, 065011 (2010).

\bibitem{SahaBook} A.A. Saharian, \textit{The Generalized Abel-Plana Formula
with Applications to Bessel Functions and Casimir Effect} (Yerevan State
University Publishing House, Yerevan, 2008); Report No. ICTP/2007/082;
arXiv:0708.1187.

\bibitem{Saha06Sum} A.A. Saharian, Proceedings of Science PoS(IC2006)019,
hep-th/0609093.

\bibitem{Prud86} A.P. Prudnikov, Yu.A. Brychkov, and O.I. Marichev, \textit{%
Integrals and Series} (Gordon and Breach, New York, 1986), Vol. 2.

\bibitem{Wats44} G.N. Watson, \textit{A Treatise on the Theory of Bessel
Functions} (Cambridge University Press, Cambridge, 1944).

\bibitem{Abra72} M. Abramowitz and I.A. Stegun, \textit{Handbook of
Mathematical Functions} (Dover, New York, 1972).

\bibitem{Spin08} J. Spinelly and E.R. Bezerra de Mello, JHEP \textbf{0812},
081 (2008).

\bibitem{Saha04} A.A. Saharian, Phys. Rev. D \textbf{69}, 085005 (2004).

\bibitem{Beze09} E.R. Bezerra de Mello and A. A. Saharian, JHEP  \textbf{0809%
}, 005 (2008).

\bibitem{Saha08} A.A. Saharian and M.R. Setare, Phys. Lett. B \textbf{659},
367 (2008); S. Bellucci and A.A. Saharian, Phys. Rev. D \textbf{77}, 124010
(2008); A.A. Saharian, Class. and Quantum Grav. \textbf{25}, 165012 (2008);
E.R. Bezerra de Mello and A.A. Saharian, JHEP \textbf{0812}, 081 (2008).

\bibitem{Hisc87} W.A. Hiscock, Phys. Lett. B \textbf{188}, 317 (1987);
M.E.X. Guimar\~{a}es, Phys. Lett. B \textbf{398}, 281 (1997); V.A. De
Lorenci and E.S. Moreira Jr, Phys. Lett. B \textbf{679}, 510 (2009).
\end{thebibliography}
\end{document}